\documentclass[twoside]{article}
\usepackage[accepted]{aistats2016}

\usepackage{times}
\usepackage{graphicx} 
\usepackage{caption}
\usepackage{subcaption}

\usepackage{amsmath}
\usepackage{amssymb}
\usepackage{enumitem}
\usepackage[usenames,dvipsnames]{xcolor}

\usepackage[round]{natbib}

\usepackage{algorithm}
\usepackage{algorithmic}

\usepackage[bookmarks=false,colorlinks=true,citecolor=Blue,linkcolor=Blue]{hyperref}

\newtheorem{theorem}{Theorem}

\newcommand{\cd}{\overset{d}{\longrightarrow}}    
\newcommand{\cas}{\overset{as}{\longrightarrow}}  
\newcommand{\tr}{^\mathrm{T}}  

\def\argmax{\operatornamewithlimits{arg\,max}}
\def\argmin{\operatornamewithlimits{arg\,min}}

\begin{document}

%
\runningtitle{Score Permutation Based  Finite Sample Inference for GARCH Models}

%
\runningauthor{B.\ Cs.\ Cs\'aji}

\twocolumn[

\aistatstitle{Score Permutation Based  Finite Sample Inference for\\
Generalized AutoRegressive Conditional Heteroskedasticity (GARCH) Models}
\aistatsauthor{Bal\'azs Csan\'ad Cs\'aji\vspace{0.5mm}}
\aistatsaddress{Institute for Computer Science and Control (SZTAKI)\\Hungarian Academy of Sciences (MTA)}]

\begin{abstract}
A standard model of (conditional) heteroscedasticity, i.e., the phenomenon that the variance of a process changes over time, is the Generalized AutoRegressive Conditional Heteroskedasticity (GARCH) model, which is especially important for economics and finance. GARCH models are typically estimated by the Quasi-Maximum Likelihood (QML) method, which works under mild statistical assumptions. Here, we suggest a finite sample approach, called ScoPe, to construct distribution-free confidence regions around the QML estimate, which have exact coverage probabilities, despite no additional assumptions about moments are made. ScoPe is inspired by the recently developed Sign-Perturbed Sums (SPS) method, which however cannot be applied in the  GARCH case. ScoPe works by perturbing the score function using randomly permuted residuals. This produces alternative samples which lead to exact confidence regions. Experiments on simulated and stock market data are also presented, and ScoPe is compared with the asymptotic theory and bootstrap approaches.
\end{abstract}

\section{INTRODUCTION}
\label{SectionIntroduction}
{\em Homoscedastic} noises, such as i.i.d.\ variables, are widely used in  learning theory \citep{vapnik1998statistical,hastie2009elements}, though most real-world phenomena, from social systems and stock markets to telecommunications and ECG signals, can be better described by {\em heteroscedastic} models as their variances change over time. It is typical that larger disturbances are more likely followed by larger disturbances, while smaller fluctuations tend to be followed by smaller fluctuations. In finance, for example, this phenomenon is called {\em volatility clustering} and it is a widely-known feature of financial time series \citep{francq2011garch}.

{\em AutoRegressive Conditional Heteroscedasticity} (ARCH) processes are standard models of this phenomenon. They were introduced by Robert F.~\citet{engle1982autoregressive} for which he was awarded the Nobel Prize for Economics in 2003 (jointly with Clive W.~J.~Granger) ``{\em for methods of analyzing economic time series with time-varying volatility (ARCH)}''. The ARCH model was later exetended by \citet{bollerslev1986generalized} who introduced its {\em generalized} version, called GARCH (see Section \ref{sec-garch}). Since then, various other generalizations have also been proposed, but GARCH is still one of the most widely used models \citep{hansen2005forecast}.

An essential question about GARCH models is how to fit them to available data. Several approaches were proposed for this, but in practice, GARCH models are almost exclusively estimated by the {\em Quasi-Maximum Likelihood} (QML) method (see Section \ref{sec-qmle}), which maximizes a (conditional) Gaussian log-likelihood function \citep{Berkes2003garch}. QML works well for a wide range of noise terms and has favorable theoretical properties, e.g., it is {\em strongly consistent} and, assuming its driving noise has finite 4th moment, {\em asymptotically normal} (see Theorem \ref{theorem-qmle} in Section \ref{sec-asym-qmle}).

The QML method provides {\em point estimates}, i.e., single models which best fit the data. Nonetheless, many applications require {\em confidence regions}, as well, showing how reliable the estimates are. They are fundamental, e.g., for risk management and robust control. The standard way of building confidence sets around a point estimate 
is to use the level sets of its limiting distribution, which typically leads to {\em asymptotic} confidence ellipsoids \citep{Ljung1999}.

This however only provides {\em approximate} confidence regions for finite datasets. Furthermore, in several cases, the noises are {\em heavy-tailed} and fail to have 4th moments, which means the asymptotic normality of the Quasi-Maximum Likelihood Estimate (QMLE) is not guaranteed. In case of relatively heavy-tailed innovations, which are common in finance, directly estimating the asymptotic distribution of QMLE becomes very difficult \citep{hall2003inference}.

Because of these issues, some authors suggested moving away from the QMLE and using other methods \citep{chan2007interval}, like the Hill estimator \citep{hill1975simple}, or applying the ML theory with other specific (non-Gaussian) distributions. The main issues with such approaches are not only that specifying a distribution introduces the risk of misspecification \citep{spierdijk2014confidence}, but also that confidence regions are typically built around a selected point estimate, and as QMLE is the most widely applied method in practice, it would be important to build confidence regions for them.

Recently there has been an increase of interest in {\em bootstrap} \citep{Efron1993} approaches particularly since some of them can create confidence regions around the QMLE even if the innovations are heavy-tailed. One of the most popular bootstrap methods for GARCH processes is the  {\em residual bootstrap} \citep{pascual2006bootstrap,shimizu2009bootstrapping} which is based on resampling (with replacement) the innovations from the {\em empirical distribution function} of the (standardized) QMLE residuals, simulating alternative trajectories based on which alternative QMLEs can be constructed. Then, typically {\em asymptotic} statistics are estimated, e.g., based on the sample of bootstrap QMLEs.

Nevertheless, standard bootstrap approaches are generally not consistent if the distribution of the asymptotic statistic is {\em non-Gaussian} and have inaccurate coverage probabilities if the true innovations are {\em skewed} \citep{hall2003inference}.

Alternatives bootstrap variants were also proposed. For example, the {\em likelihood ratio} (LR) bootstrap \citep{luger2012finite} for {\em stationary} GARCH processes is based on defining a $p$-value using conventional LR hypothesis tests combined with bootstrap. For a particular parameter it builds alternative bootstrap trajectories and computes bootstrap LRs. These are then compared with the LR of the original parameter. This approach can lead to {\em finite sample} guarantees, but only for completely {\em known} noise distributions, moreover, it is computationally demanding as it involves computing several bootsrap ML estimates for each parameter it tests.

Here, we propose a {\em finite sample} inference technique for GARCH models which can be seen as (i) an {\em exact} hypothesis test as well as (ii) a way to construct {\em distribution-free}, exact confidence regions around the QMLE without additional statistical assumptions about moments or stationarity. Its core is a permutation type test \citep{good2005permutation} and it is called {\em ScoPe} as it applies randomly {\em permuted}  residuals in the {\em score} function, i.e., the gradient of the log-likelihood (see Section \ref{sec-scope}).
ScoPe was inspired by the recently developed {\em Sign-Perturbed Sums} (SPS) identification 
algorithm \citep{Csaji2012b, Csaji2014, Csaji2015}, which can build exact, non-asymptotic, distribution-free confidence regions around the prediction error estimate of general linear dynamical systems; however, SPS cannot be applied for GARCH models.

We should note that an exact, distribution-free permutation test in the context of GARCH models was also proposed by \citet{luger2012testing,luger2014testing}. However, that test permutes the GARCH process itself (not the residuals in the score function) and it is {\em only} applicable to test the hypothesis of {\em conditional homoskedasticity}. Particularly, it cannot be used to build confidence regions for the GARCH parameters.

\section{GARCH MODELS}
\label{sec-garch}
Formally, a GARCH($p$, $q$) process, $\{X_t\}$, is defined by the following two equations \citep{francq2011garch}
\begin{subequations}
\begin{eqnarray}
X_t & \!\!\triangleq\!\! & \sigma_t \,\varepsilon_t,\label{GARCH1}\\
\sigma^2_t & \!\!\triangleq\!\! & \omega^* + \,\sum_{i=1}^p \alpha_i^* X^2_{t-i}\,+\,\sum_{j=1}^q \beta_j^* \sigma^2_{t-j}\label{GARCH2},
\end{eqnarray}
\end{subequations}
where $\{\varepsilon_t\}$ is a strong white noise, i.e., an i.i.d.\ sequence of real random variables with zero mean and unit variance;
variable $\sigma^2_t$ is latent and 
defines the conditional variance of $X_t$, given its own past up to $t-1$; and $\omega^* > 0 $ as well as $\alpha_i^*, \beta_j^* \geq 0$ are constants, where $1 \leq i \leq p$ and $1 \leq j \leq q$. Integers $p$ and $q$ are called the {\em orders} of the model. In case $q=0$, we get back Engle's classical ARCH model.

It is known \citep{bollerslev1986generalized} that there exists a wide-sense stationary solution to (\ref{GARCH1})-(\ref{GARCH2}) if and only if 
\begin{equation}
\label{stationary}
\sum_{i=1}^p \alpha_i^* + \sum_{j=1}^q \beta_j^* < 1.
\end{equation}
If $\{X_t\}$ is a wide-sense stationary GARCH process, it is necessarily also strictly stationary. 
Moreover, it is a (potentially scaled) weak white noise, that is $\mathbb{E}[X_t] = 0$, $\mathbb{E}[X_tX_k]=0$, and $\mathbb{E}[X^2_t] = \eta$,
for all $t$ and  $k \neq t$, where $\mathbb{E}[\cdot]$ denotes expectation and $\eta$ can be calculated by
\begin{equation*}
 \eta \, =\, \frac{\omega^*}{1 - \sum_{i=1}^q\alpha^*_i - \sum_{j=1}^p \beta^*_j}.
\end{equation*}

Conditions for the {\em unique} existence of a strictly stationary and causal solution to system (\ref{GARCH1})-(\ref{GARCH2}) can be given in terms of the top Lyapunov exponent of its (Markovian) state space representation \citep{Straumann2005}.

\section{QUASI-MAXIMUM LIKELIHOOD}
\label{sec-qmle}
While there are several approaches to estimate GARCH processes, such as prediction error methods or the Whittle estimator \citep{Straumann2005}, the most widely used estimators belong to the class of {\em Quasi-Maximum Likelihood} (QML) methods \citep{Berkes2003garch}. They are typically applied off-line, while there is also a recursive extension of the QML theory \citep{gerencser2012real}. Now, we briefly recall the QML method for GARCH processes.

We do not make further assumptions on the distribution of the noise terms, $\{ \varepsilon_t \}$, but accept a ``working hypothesis'' that they are Gaussian. This however will not be needed to apply the method: as we will see, the QML method works well under very mild statistical assumptions. 

More precisely, if we were to assume that $\{ \varepsilon_t \}$ were Gaussian (hence standard normal, since we assumed that $\mathbb{E}[\varepsilon_t] = 0$ and $\mathbb{E}[\varepsilon^2_t] = 1$), then the conditional distribution of $X_t / \sigma_t$, given the $\sigma$-algebra generated by $\{ \varepsilon_k \}_{k<t}$, would also be standard normal. The {\em quasi maximum likelihood estimate} (QMLE) is derived under this hypothesis.

Assuming known initial values $X_0(\theta), \dots, X_{1-p}(\theta)$ and $\widehat{\sigma}_0(\theta), \dots, \widehat{\sigma}_{1-q}(\theta)$, to be discussed below, the {\em
conditional Gaussian quasi-likelihood} function is defined as
\begin{equation*}
\mathcal{L}_n(\theta) = \mathcal{L}_n(\theta; x) \, \triangleq \,\prod_{t=1}^n \frac{1}{\sqrt{2 \pi \widehat{\sigma}^2_t(\theta)}}\exp\left(-\frac{X_t^2}{2 \widehat{\sigma}_t^2(\theta)}\right),
\end{equation*}
where $x = (X_1, \dots, X_n)$ is the available sample and 
\begin{equation}
\label{condvarest}
\widehat{\sigma}_t^2(\theta) \, \triangleq \,  \omega + \sum_{i=1}^p \alpha_i X^2_{t-i}+\sum_{j=1}^q \beta_j \widehat{\sigma}^2_{t-j}(\theta),
\end{equation}
where $\theta \in \mathbb{R}^{p+q+1}$ is a generic vector encoding the parameters,
$\theta \triangleq  (\omega, \alpha_1, \dots, \alpha_p, \beta_1, \dots, \beta_q),$
while the ``true'' parameter vector is denoted by 
$\theta^* \triangleq (\omega^*, \alpha_1^*, \dots, \alpha_p^*, \beta_1^*, \dots, \beta_q^*)$.

We need initial conditions to calculate $\widehat{\sigma}_t^2(\theta)$ recursively. Standard choices include zero or
$X^2_0(\theta) = \dots = X^2_{1-p}(\theta) = \widehat{\sigma}_0^2(\theta) = \dots = \widehat{\sigma}_{1-q}^2(\theta) = \omega$,
or the unconditional variance w.r.t.\ $\theta$ \citep{francq2011garch}
\begin{equation*}
X^2_0(\theta) = \dots = X^2_{1-p}(\theta) = \widehat{\sigma}_0^2(\theta) = \dots = \widehat{\sigma}_{1-q}^2(\theta) =
\end{equation*}
\begin{equation*}
\frac{\omega}{1 - \sum_{i=1}^q\alpha_i - \sum_{j=1}^p \beta_j}.
\end{equation*}

The QMLE is any measurable solution of the problem
\begin{equation*}
\widehat{\theta}_n \,\triangleq \, \argmax_{\theta \in \Theta} \mathcal{L}_n(\theta),
\end{equation*}
where $\Theta$ is the set of allowed parameters, for example $\theta \in \Theta$ if $\omega > 0 $, $\alpha_i^*, \beta_j^* \geq 0$, for all $i,j$, and there exists a stationary solution to (\ref{GARCH1})-(\ref{GARCH2}), i.e., property (\ref{stationary}) holds.

Taking the natural logarithm, $\log(\cdot)$, of $\mathcal{L}_n(\theta)$ leads to the conditional quasi-log-likelihood function
\begin{equation*}
\ell^*_n(\theta) = \ell^*_n(\theta; x)\,\triangleq\, \log \mathcal{L}_n(\theta) = \log \mathcal{L}_n(\theta; x),
\end{equation*}
which, under the standard normal hypothesis, simplifies to
\begin{equation*}
\ell^{*}_n(\theta) = -\frac{1}{2} \sum_{t=1}^n \left[\,\log(2 \pi) + \log \widehat{\sigma}_t^2(\theta) + \frac{X_t^2}{\widehat{\sigma}_t^2(\theta)}\,\right].
\end{equation*}
Since the optimal point is not affected by the constants,
maximizing $L_n(\theta)$ is equivalent to minimizing $\ell_n(\theta)$,
\begin{equation*}
\ell_n(\theta) \, \triangleq \, \frac{1}{n} \sum_{t=1}^n \left[\,\log \widehat{\sigma}_t^2(\theta) + \frac{X_t^2}{\widehat{\sigma}_t^2(\theta)}\,\right],
\end{equation*}
where the $1/n$ term is included for numerical stability. The minimization of $\ell_n(\theta)$ is typically done by an iterative numerical method, such as the Newton-Raphson algorithm.

\section{ASYMPTOTICS OF QMLE}
\label{sec-asym-qmle}
It is known that QMLE is strongly consistent and asymptotically normal. In order to make these claims more precise, let  us introduce some assumptions \citep{Straumann2005}. When we talk about the marginal distribution of a stationary process $\{Y_t\}$, a generic element will be denoted by $Y_0$.
\begin{enumerate}[label=(Q\arabic*)]
\renewcommand{\theenumi}{Q\arabic{enumi}}
\item\label{Assume-Not-Concentrated}{\em The noise is nondegenerate: the distribution of $\varepsilon_0$ is not concentrated in two points.}
\item\label{Assume-Identifiable}{\em The process is identifiable, that is $(\alpha^*_p, \beta^*_q) \neq (0, 0)$, $\omega > 0$, 
$\exists \, i:\alpha^*_i > 0$, and the polynomials $p(z) \triangleq \alpha_1^* z + \dots + \alpha_p^* z^p$ and $q(z) \triangleq \beta_1^* z + \dots + \beta_q^* z^q$ do not have any common zeros.}
\item\label{Assume-Interior}{\em The true parameter, $\theta^*$, is in the interior of $\Theta$.}
\item\label{Assume-Not-Too-Mass-Around-Zero}{\em } 
$\exists \, \mu > 0$, such that $\mathbb{P}(|\varepsilon_0| \leq t) = o(t^\mu)$ as $t\downarrow 0$.
\end{enumerate}

Then, using the four assumptions above, it can be proven that \citep{Berkes2003garch,Straumann2005}

\begin{theorem}
\label{theorem-qmle} Let $\{X_t\}$ be a stationary GARCH$(p,q)$ process with true parameter $\theta^* \in \Theta$. Then, assuming \ref{Assume-Not-Concentrated} and \ref{Assume-Identifiable}, QMLE is strongly consistent, that is
\begin{equation*}
\widehat{\theta}_n \cas \theta^* \hspace{5mm} \mbox{as} \hspace{5mm} n \to \infty.
\end{equation*}
If additionally $\mathbb{E}[\varepsilon^4_0] < \infty$ and \ref{Assume-Interior}, \ref{Assume-Not-Too-Mass-Around-Zero} hold, the QMLE is also asymptotically normally distributed, i.e.
\begin{equation*}
\sqrt{n}(\widehat{\theta}_n - \theta^*) \cd \mathcal{N}(0, F_0^{-1} G_0 F_0^{-1})  \hspace{5mm} \mbox{as} \hspace{5mm} n \to \infty,
\end{equation*}
where $F_0, G_0$ are $(p\times q\times 1) \times (p\times q\times 1)$ matrices 
\begin{eqnarray*}
F_0 &\!=\!& -\frac{2}{\mathbb{E}\,[\varepsilon^4_0-1]}\,G_0,\\[5pt]
G_0 &\!=\!& \frac{\mathbb{E}\,[\varepsilon^4_0-1]}{4}\,\, \mathbb{E}\!\left[\frac{1}{\sigma_0^4} \nabla_{\!\theta}\, \widehat{\sigma}_0^2(\theta^*) \nabla_{\!\theta}\, \widehat{\sigma}^2_0(\theta^*)\tr \right],
\end{eqnarray*}
where $\mathcal{N}(m, C)$ denotes the (multivariate) Gaussian distribution with mean vector $m$ and covariance matrix $C$, while $\nabla_{\!\theta}f(\cdot)$ is the gradient vector of $f(\cdot)$ with respect to $\theta$.
\end{theorem}

In many applications not just a point estimate, like QMLE, but a confidence region is also needed. The standard approach in practice is to use the (level sets of the) asymptotic distribution of the estimate to build a confidence region \citep{Ljung1999,Soderstrom1989}. 

To be more specific, assume we have an estimate $\Gamma_n$ of the true covariance matrix $F_0^{-1} G_0 F_0^{-1}$, based on $n$ data points. Then, an
asymptotic confidence ellipsoid can be built by
\begin{equation}
\label{asymptotic ellipsoid}
\widetilde{\Theta}_{n}(s) \triangleq  \bigg\{\, \theta \in \mathbb{R}^{d} : (\theta - {\widehat{\theta}_n})^\mathrm{T}\, \Gamma^{-1}_n\, (\theta - {\widehat{\theta}_n}) \, \leq \, \frac{s}{n} \,\bigg\},
\end{equation}
where $d \triangleq p+q+1$ and the 
probability that the true parameter is covered, i.e., $\theta^* \in \widetilde{\Theta}_n$, is {\em approximately} $F_{\chi^2(d)}(s)$, where $F_{\chi^2(d)}$ is the cumulative distribution function (CDF) of the standard $\chi^2$ distribution with $d$ degrees of freedom. 

The main problems with such an approach are that (i) even if the covariance matrix of the asymptotic distribution was known exactly, the levels sets for a finite sample would still be only {\em approximately} correct, as they rely on a result which is only guaranteed in the limit.
Moreover, (ii) the asymptotic normality of the estimation error requires finite 4th moment from the noise, 
which is often not the case in practice, e.g., for heavy-tailed distributions.
Hence, such ellipsoids can only be used as {\em heuristics} in a finite sample setup, as their confidence levels are not guaranteed.

\section{FINITE SAMPLE, DISTRIBUTION- FREE CONFIDENCE REGIONS}
\label{sec-sps}
Now we turn our attention to finite sample, distribution-free results to overcome the issues mentioned above. In the next section, the ScoPe method is introduced to construct non-asymptotic, distribution-free confidence regions around the QML estimate for GARCH processes, which have {\em exact} confidence probabilities. The main motivation of the suggested approach comes from the {\em Sign-Perturbed Sums} (SPS) algorithm \citep{Csaji2012b, Csaji2014, Csaji2015}. 
Though, SPS cannot be used for GARCH processes, for reasons discussed below, we briefly present it here to motivate our permutation-rank based method.

Let us consider the following scalar general linear dynamical system \citep{Ljung1999,BoxJenkins2008}:
\begin{equation*}
\label{genlin}
Y_t \, \triangleq \, G(z^{-1}; \theta^*)\, U_t + H(z^{-1}; \theta^*)\, N_t,
\end{equation*}
where $t$ denotes (discrete) time, $Y_t$ is output, $U_t$ is an input, $N_t$ is a noise, 
$G$, $H$ are (causal) rational transfer functions, and $z^{-1}$ is the backward shift operator. As previously, $\theta^*$ denotes the unknown true parameter of the system. The assumptions on the system are as follows \citep{Csaji2012b}
\begin{enumerate}[label=(S\arabic*)]
\renewcommand{\theenumi}{S\arabic{enumi}}
\item\label{S1}{\em The ``true'' system is in the 
model class which has polynomials with known orders.}
\item\label{S2}{\em $H(\,z^{-1}; \theta\,)$ has a stable inverse, $G(\,0\,; \theta\,) = 0$ and $H(\,0\,; \theta\,) = 1$\,,\, for\, $\theta \in \Theta$.}
\item\label{S3}{\em The noise $\{N_t\}$ is independent, and each $N_t$ has a symmetric distribution about zero.}
\item\label{S4}{\em The system operates in open-loop, i.e., the inputs $\{U_t\}$ are independent of $\{N_t\}$}.
\item\label{S5}{\em The initialization of the system is known; for simplicity, we use \,$Y_t = N_t = U_t = 0$\,,\, for\, $t \leq 0$}.
\end{enumerate}

Under these assumptions, the noise terms can be reconstructed, given a particular $\theta$, by
\begin{equation*}
\label{linear-noise-reconstruction}
\widehat{N}_t(\theta) \, \triangleq \, H^{-1}(z^{-1}; \theta)(Y_t - G(z^{-1}; \theta)\, U_t ),
\end{equation*}
which are called
{\em residuals} or {\em prediction errors}. It is important to note that $\widehat{N}_t(\theta^*) = N_t$, for all $t$. 

The {\em prediction error estimate}, $\tilde{\theta}_n$, is defined as the minimizer of the squared prediction errors \citep{Ljung1999},
\begin{equation*}
\tilde{\theta}_n\,\triangleq\, \argmin_{\theta \in \Theta} \sum_{t=1}^n\widehat{N}_t^2(\theta),
\end{equation*}
which can be found be solving the {\em normal equation},
\begin{equation*}
\sum_{t=1}^n \widehat{N}_t(\tilde{\theta}_n)\nabla_{\!\theta}\widehat{N}_t(\tilde{\theta}_n)\,=\,0.
\end{equation*}
where $\Theta$ contains the allowed models, e.g., stable systems.

SPS builds its confidence region by perturbing the normal equation: 
given a 
$\theta$, 
it builds $m-1$ alternative output trajectories using perturbed versions of the estimated residuals,
\begin{equation*}
\bar{Y}_{t}(\theta, \alpha_i) \, \triangleq \, G(z^{-1}; \theta)\, U_t + H(z^{-1}; \theta)\,(\alpha_{i,t} \widehat{N}_t(\theta)),
\end{equation*}
where $\{\alpha_{i,t}\}$ are $(m-1) \times n$ i.i.d.\ random signs, that is random variables which take values $\pm 1$ with probability $1/2$ each; and $\alpha_i$ denotes the vector $(\alpha_{i,1}, \dots, \alpha_{i,n})$. Note that $n$ is the sample size of the residuals we can reconstruct from $\{Y_t\}$,
and $m$ is a user-chosen design parameter.

Let us denote $\nabla_{\!\theta}\widehat{N}_t(\theta)$ by $\psi_t(\theta)$. Then,
$\psi_t(\theta)$ can be treated as a linear filter on $\{Y_t\}$ and $\{U_t\}$,
\begin{equation*}
\psi_t(\theta)\,= \, W_0(z^{-1};\theta)\,Y_t + W_1(z^{-1};\theta)\,U_t,
\end{equation*}
where $W_0$ and $W_1$ are vector-valued linear filters \citep{Ljung1999}. We produce perturbed versions of $\psi_t(\theta)$ by
\begin{equation*}
\bar{\psi}_t(\theta, \alpha_i)\,\triangleq\,W_0(z^{-1};\theta)\,\bar{Y}_{t}(\theta, \alpha_i) + W_1(z^{-1};\theta)\,U_t,
\end{equation*}
where $i \in \{1, \dots, m\}$, and define a reference function, $S_0$, and\, $m-1$ sign-perturbed functions, $\{S_i\}$, as
\begin{eqnarray*}
S_0(\theta)\!\! & \triangleq\!\! & \Psi_n^{-\frac{1}{2}}(\theta) \sum_{t=1}^{n}{\, \psi_t(\theta) \widehat{N}_t(\theta)},\\[5pt]
S_i(\theta) \!\! & \triangleq\!\! & \bar{\Psi}_n^{-\frac{1}{2}}(\theta,\alpha_i) \sum_{t=1}^{n}{\, \alpha_{i,t} \,
\bar{\psi}_t(\theta,\alpha_i) \widehat{N}_t(\theta)},
\end{eqnarray*}
where 
$\Psi_n$ and $\bar{\Psi}_n(\theta, \alpha_i)$ are covariance estimates, only used to shape the confidence region \citep{Csaji2012b}.

Let use denote by $\mathcal{R}^0_m(\theta)$ the position of $\|S_0(\theta)\|^2$ in the ordering of variables $\{\|S_i(\theta)\|^2\}$, where ties are broken randomly. Therefore, $\mathcal{R}^0_m(\theta) = 1$ if $\|S_0(\theta)\|^2$ is the smallest in the ordering, $\mathcal{R}^0_m(\theta) = 2$ if it is the second smallest and so on. Then, the SPS confidence region is built by
\begin{equation*}
\widetilde{\Theta}_n(m,r) \, \triangleq \, \left\{\, \theta \in \Theta\, :\, \mathcal{R}^0_m(\theta) \leq m-r\, \right\},
\end{equation*}
where $m > r > 0$ are user-chosen integers. The SPS region has exact confidence probability \citep{Csaji2012b}:
\vspace{-2mm}
\begin{theorem} Under assumptions \ref{S1}, \ref{S2}, \ref{S3}, \ref{S4}, \ref{S5}, 
$$\mathbb{P}\bigl(\,\theta^* \in \widetilde{\Theta}_n(m,r)\,\bigr)\, =\, 1 - \frac{r}{m}.$$
\end{theorem}
Since $\|S_0(\tilde{\theta}_n)\|^2 = 0$, i.e.,  the prediction error estimate satisfies the normal equation, it is always in the confidence region, assuming it is non-empty. In the special case of linear regression problems in which the regressors are independent of the noise, for example, generalized finite impulse response systems, it can be proved, as well, that the SPS confidence region is strongly consistent \citep{Csaji2014} and it is also star convex with the least-squares estimate as a star center \citep{Csaji2015}. Nevertheless, the main strength of SPS lies in the fact that its confidence probability is {\em exact}, i.e., its confidence sets are non-conservative.

\section{SCORE PERMUTATION}
\label{sec-scope}
In this section, inspired by the core ideas underlying SPS,
we present the {\em Score Permutation} (ScoPe) method, which can
construct {\em exact}, {\em finite sample}, {\em distribution-free} confidence regions around the QMLE of GARCH processes.

Unfortunately, SPS cannot be 
applied to build such confidence regions,
since, e.g., (i) SPS is defined for linear systems, while GARCH processes are nonlinear; (ii) it is built for a quadratic cost criterion, 
not for the QMLE; moreover, (iii) the GARCH residuals appear squared in the dynamics of the conditional variances, see (\ref{GARCH2}), thus, perturbing their signs does not produce alternative variance trajectories. Instead of sign-perturbations, 
ScoPe uses random permutations in the spirit of statistical permutation-rank 
tests \citep{good2005permutation}. A random permutation matrix based alternative to SPS was also analyzed by \citet{kolumban2015perturbed} for linear regression problems with deterministic regressors.

Recall that the QMLE satisfies the {\em likelihood equation},
\begin{equation*}
\nabla_{\!\theta}\,\ell_n(\widehat{\theta}_n ) = 0,
\end{equation*}
and the gradient of the (conditional) log-likelihood function, the {\em score} function, can be written as
\begin{equation*}
\nabla_{\!\theta}\,\ell_n(\theta ) = \frac{1}{n} \sum_{t=1}^n \left(1 - \frac{X_t^2}{\widehat{\sigma}_t^2(\theta)}\right) \frac{1}{\widehat{\sigma}_t^2(\theta)} \nabla_{\!\theta}\,\widehat{\sigma}_t^2(\theta) =
\end{equation*}
\begin{equation*}
\frac{1}{n} \sum_{t=1}^n \left(1 - \widehat{\varepsilon}^{\,2}_t(\theta)\right) \frac{1}{\widehat{\sigma}_t^2(\theta)} \nabla_{\!\theta}\,\widehat{\sigma}_t^2(\theta),
\end{equation*}
where $\widehat{\varepsilon}_t(\theta) \triangleq X_t/\widehat{\sigma}_t(\theta)$ is a reconstructed residual (innovation) for time $t$ assuming a particular parameter $\theta$, and $\widehat{\sigma}_t(\theta)$ is an estimate of $\sigma_t$, which can be calculated recursively using (the square root of) formula (\ref{condvarest}).

We can observe that $\widehat{\varepsilon}_t(\theta^*) = \varepsilon_t$, for all $t$, assuming the initial conditions 
are known, more precisely
\vspace{-1mm}
\begin{enumerate}[label=(P\arabic*)]
\renewcommand{\theenumi}{P\arabic{enumi}}
\item\label{P1}{\em The ``true'' system is in the model class, i.e., upper bounds on the orders $p,q$ are known and $\theta^*$ is in $\Theta$.}
\item\label{P2}{\em The initial conditions 
$\widehat{\sigma}^2_0(\theta), \dots, \widehat{\sigma}^2_{1-q}(\theta)$ of the conditional variances are known, i.e.,
$\widehat{\sigma}^2_t(\theta^*) = \sigma^2_t$.}
\end{enumerate}
\vspace{-1mm}
Initial conditions for $\{X^2_t\}$ are not needed, as system (\ref{GARCH2}) is autoregressive with finite order  in the $X^2_t$ variables and $\{X_t\}$ is observed. Henceforth, for notational simplicity, we assume (w.l.o.g.) that $X_0, \dots, X_{1-p}$ are available.

Note that \ref{P1} is a standard assumption and it is even needed to define the concept of confidence regions (namely, a subset of parameters which contains the ``true'' parameter with at least a given probability). Assumption \ref{P2} is also typical, especially for methods aiming at finite sample guarantees. It is mild, as it can be simply omitted if the system is ARCH (which was Engle's original model). Even if it is GARCH, the autocorrelation of conditional variances decays exponentially, therefore, it is expected that the effect of violating this assumption vanishes as we have more and more data. This is also supported by our experiments (Section \ref{sec-exp}).

Since $\{\varepsilon_t\}$ is i.i.d., its every permutation 
results in a sequence having the same distribution, namely
\begin{equation*}
\{\varepsilon_t\}\,\,{\buildrel d \over =}\,\,\{\varepsilon_{\pi(t)}\}
\end{equation*}
where $\pi(\cdot)$ is an arbitrary permutation on the indices, i.e., a bijection of $\{1, \dots, n\}$ onto itself.

Given a parameter $\theta$, the main idea is to first ``invert'' the system to get the residuals $\{\widehat{\varepsilon}_t(\theta)\}$ and then generate 
alternative trajectories by applying random permutations on their indices. More precisely, we must generate $m-1$ random permutations $\pi_1, \dots, \pi_{m-1}$, where each permutation has the same probability $1/n!$ to be selected. Then, we can define alternative residuals for parameter $\theta$ by
\begin{equation*}
\widehat{\varepsilon}_{\pi_i(1)}(\theta), \dots, \widehat{\varepsilon}_{\pi_i(n)}(\theta),
\end{equation*}
for all $i \in \{1, \dots, m-1\}$, where $m$ is a user-chosen parameter as before. For simplicity, we denote the identity permutation by $\pi_0$, i.e., $\pi_0(t) = t$, for all $t$. In some cases it is useful to first standardize the residuals, by subtracting their sample mean and dividing with their standard deviation. Using this notation, the original and the perturbed score function (the gradient of the log-likelihood) is
\begin{equation*}
B(\theta, \pi_i) \, \triangleq \, \frac{1}{n} \sum_{t=1}^n \frac{\left(1 - \widehat{\varepsilon}^{\,2}_{\pi_i(t)}(\theta)\right)}{\bar{\sigma}_t^2(\theta, \pi_i)} \nabla_{\!\theta}\,\bar{\sigma}_t^2(\theta, \pi_i),
\end{equation*}
where the perturbed variances $\bar{\sigma}_t^2(\theta, \pi_i)$ are defined as
\begin{equation*}
\bar{\sigma}_t^2(\theta, \pi_i)\, \triangleq \,\omega + \sum_{k=1}^p \alpha_k \bar{X}^2_{t-k}(\theta, \pi_i) + \sum_{j=1}^q \beta_j \bar{\sigma}^2_{t-j}(\theta, \pi_i),
\end{equation*}
with the same initial values for all generated permutations,
$$
\bar{\sigma}_0^2(\theta, \pi_i) \,\triangleq \,\widehat{\sigma}_0(\theta), \dots, \bar{\sigma}_{1-q}^2(\theta, \pi_i)\, \triangleq \, \widehat{\sigma}_{1-q}(\theta).
$$
This gives rise to an alternative output trajectory with
\begin{equation}
\label{alternative-trajectory}
\bar{X}_{t}(\theta, \pi_i) \,\triangleq\, \bar{\sigma}_{t}(\theta, \pi_i) \, \widehat{\varepsilon}_{\pi_i(t)}(\theta).
\end{equation}
Observe that $B(\theta, \pi_0) = \nabla_{\!\theta}\,\ell_n(\theta )$ as $\pi_0$ is the identity permutation. We use $\|B(\theta, \pi_0)\|^2$ as a reference and compute its rank in the ordering of $\{\|B(\theta, \pi_i)\|^2\}$ variables. There is a chance two such functions take on the same value, e.g., if the noise is discrete. In order to handle this, we use a tie-breaking,
namely, with the help of another random permutation $\nu$. This one is on $\{0, \dots m-1\}$. Given $m$ real numbers $Z_0, \dots, Z_{m-1}$ we define a strict total order $\succ_{\nu}$ as 
$$Z_k \succ_{\nu} Z_j \hspace{10mm}\hbox{if and only if}$$
$$\left(\,Z_k > Z_j\,\right) \hspace{2mm}\hbox{or}\hspace{2mm} \left(\,Z_k = Z_j \hspace{2mm}\hbox{and}\hspace{2mm} \nu(k) > \nu(j)\,\right).$$
The {\em rank} of $\|B(\theta, \pi_0)\|^2$ w.r.t.~the ordering $\succ_{\nu}$ is then 
$$
\mathcal{R}_m(\theta) \,\triangleq\, 1 + \sum_{i=1}^{m-1} \mathbb{I}\!\left( \|B(\theta, \pi_0)\|^2 \succ_{\nu} \|B(\theta, \pi_i)\|^2 \right),
$$
where $\mathbb{I}(\cdot)$ is an indicator function: 
it is $1$ if its argument is true and $0$ otherwise. The ScoPe confidence set is
\begin{equation*}
\widehat{\Theta}_n(m,r) \, \triangleq \, \left\{\, \theta \in \Theta\, :\, \mathcal{R}_m(\theta) \leq m-r\, \right\},
\end{equation*}
where $m > r > 0$ are user-chosen integers affecting the coverage probability of the confidence region. The main theoretical claim of the paper is the following theorem.

\begin{theorem}\label{mainthm}Assuming \ref{P1} and \ref{P2}, we have \,
$$\mathbb{P}\bigl(\,\theta^* \in \widehat{\Theta}_n(m,r)\,\bigr)\, =\, 1 - \frac{r}{m}.$$
\end{theorem}
\subsubsection*{Proof of Theorem 3}
The main idea is to show that $\{\|B(\theta^*, \pi_i)\|^2\}$ are {\em conditionally} i.i.d.\ (for the case of the true parameter, $\theta^*$), therefore exchangeable, which leads to the fact that each ordering of them has the same probability, namely, $1/m!$. 

By definition,  $\theta^* \in \widehat{\Theta}_n(m,r)$ if and only if $\mathcal{R}_m(\theta^*) \leq m-r$, i.e., if $\|B(\theta^*, \pi_0)\|^2$ takes one of the positions $1, \dots, m-r$ in the ordering of $\{\|B(\theta^*, \pi_i)\|^2\}$ variables, 
with respect to $\succ_{\nu}$. Our main aim will be to prove that $\{\|B(\theta^*, \pi_i)\|^2\}$ are {\em uniformly ordered}, that is to show that 
\begin{equation*}
\mathbb{P}\left( \|B(\theta^*, \pi_{\gamma(0)})\|^2\!\succ_{\nu}\!\dots\!\succ_{\nu}\!\|B(\theta^*, \pi_{\gamma(m-1)})\|^2\right) = \frac{1}{m!},
\end{equation*}
for all possible permutation $\gamma$ on $\{0, \dots, m-1\}$. From this, the theorem follows immediately, since then $\|B(\theta^*, \pi_0)\|^2$ takes each position in the ordering with probability exactly $1/m$, thus $\mathbb{P}(\mathcal{R}_m(\theta^*) = i) = 1/m$ for $i \in \{0, \dots, m-1\}$, from which $\mathbb{P}\bigl(\theta^* \in \widehat{\Theta}_n(m,r)\bigr)\, =\, 1 - r/m$.

Before we show the uniform ordering of $\{\|B(\theta^*, \pi_i)\|^2\}$, we introduce some notations and state some useful facts about random permutations.

If $\gamma$ is a permutation on $\{1, \dots, n\}$ and $Z = (Z_1, \dots, Z_{n})$ is a vector of dimension $n$, then let
$$
\gamma(Z) \,\triangleq\, (Z_{\gamma(1)}, \dots, Z_{\gamma(n)}).
$$
The inverse of a permutation $\gamma$ is denoted by $\gamma^{-1}$, that is $\gamma^{-1}(\gamma(Z)) = Z$. If $\gamma$ and $\pi$ are permutations, their composition is denoted by $\gamma \circ \pi$, that is $(\gamma \circ \pi)(Z) = \gamma(\pi(Z))$.

It can be proven that if $Z = (Z_1, \dots, Z_{n})$ is an i.i.d. random vector and $\gamma$ is a random permutation which is uniformly chosen from all possible permutations of $\{1, \dots, n\}$, with $\gamma$ being independent of $Z$, then we can conclude that $\gamma$ and $\gamma^{-1}(Z)$ are independent, as well.

Also if $\gamma, \pi_1, \dots, \pi_{k}$ are $k+1$ i.i.d.\ uniformly chosen random permutations, then $\gamma, \pi_1 \circ \gamma^{-1}, \dots, \pi_{k} \circ \gamma^{-1}$ are also $k+1$ i.i.d.\ random permutations (also uniform).

Finally, if $Z_0, \dots, Z_{m-1}$ are i.i.d.\ random variables, then they are uniformly ordered w.r.t.~$\succ_{\nu}$ \citep[Lemma 3]{Csaji2015}. Note that this is even the case for discrete random variables, since $\succ_{\nu}$ takes care of the tie-breaking; recall that $\nu$ is a random permutation on $\{0, \dots, m-1\}$.

Now, we proceed with the proof by showing the uniform ordering property of $\{\|B(\theta^*, \pi_i)\|^2\}$ variables. 

In order to simplify the notations, let us introduce
$$
f(\varepsilon, \pi_i) \,\,\triangleq\,\, \|B(\theta^*, \pi_i)\|^2,
$$
for indices $i \in \{0, \dots, m-1\}$, where $\varepsilon = (\varepsilon_1, \dots, \varepsilon_n) = (\widehat{\varepsilon}_1(\theta^*), \dots, \widehat{\varepsilon}_n(\theta^*))$.  Note that here we used our assumptions 
\ref{P1} and \ref{P2},
i.e., that we could reconstruct the ``true'' noise sequence $\varepsilon$, in case we knew the true parameter $\theta^*$.

Let us introduce a new (uniform) random permutation $\mu$ on $\{1, \dots, n\}$, generated independently of $\pi_1, \dots, \pi_{m-1}$. We can ``inject'' the new permutation $\mu$ into our system by
$$
f(\mu(\varepsilon), \pi_i \circ \mu^{-1}) \,=\, f(\varepsilon, \pi_i) ,
$$
for all $i \in \{0, \dots, m-1\}$, since we simply undo the effect of permutation $\mu$ by composing $\pi_i$ with its inverse.

Now, let us fix a realization of $\mu(\varepsilon)$, denoted by $r$, i.e., from now on we condition on this realization. Then, the only random element in the variables $W_i \triangleq f(r, \pi_i \circ \mu^{-1})$ are $\pi_i \circ \mu^{-1}$. Now, we know that $\pi_0$ is the identity permutation, but with injecting $\mu$ we have managed to ``re-randomize'' it without chaining the method. By using the previously mentioned fact, we know that $\mu^{-1}, \pi_1 \circ \mu^{-1}, \dots, \pi_{m-1} \circ \mu^{-1}$ are i.i.d.~random elements. Since applying the same function to elements of an i.i.d.~collection results in an i.i.d.\ collection, it follows that $W_0, \dots, W_{m-1}$ are i.i.d. Hence, they are uniformly ordered w.r.t.~$\succ_{\nu}$, {\em conditionally on $r$}.

Until now, we have showed that $\{\|B(\theta^*, \pi_i)\|^2\}$ are uniformly ordered given a realization of $\mu(\varepsilon)$. To get rid of this conditioning, we can observe that (i) the ordering distribution we got is independent of the actual realization of $\mu(\varepsilon)$, and (ii) vector $\mu(\varepsilon)$ is independent of $\mu^{-1}, \pi_1 \circ \mu^{-1}, \dots, \pi_{m-1} \circ \mu^{-1}$. In this case, we know \citep[Lemma 2]{Csaji2015} that the uniform ordering also holds without conditioning on a realization. $\Box$

\vspace{1mm}
Now, let us make some remarks on ScoPe:
\vspace{-2mm}
\begin{itemize}
\item[(i)] The confidence probability is {\em exact} for any {\em finite sample}, thus no conservativism is introduced.
\item[(ii)] Parameters $m$ and $r$ are user-chosen, hence, the confidence probability is {\em under our control}.
\item[(iii)] The applied statistical assumptions are very mild, e.g., we do not assume knowing the particular distribution of the noise, i.e., it is a {\em distribution-free} method.
\item[(iv)] Unlike the standard asymptotic ellipsoids or bootstrap approaches, the confidence probability of ScoPe is exact even for {\em heavy-tailed} and {\em skewed} distributions.
\item[(v)] Scope neither needs assumptions on stationary, therefore, it can work for {\em nonstationary} processes.
\item[(vi)] Since the QMLE satisfies the likelihood equation, i.e., $\nabla_{\!\theta}\,\ell_n(\widehat{\theta}_n) = 0$, we have $\|B(\widehat{\theta}_n , \pi_0)\|^2 = 0$. Hence, the {\em QMLE is always included} in the confidence set, assuming it is non-empty. In other words, ScoPe builds its confidence regions around the QML estimate.
\item[(vii)]Finally, if we evaluate $\mathcal{R}_m(\theta)$ on a set of $\theta$ values, their ranks indicate which confidence levels can be associated to those parameters, therefore, these ``rank fields'' contain information about the distribution of the estimation error. On the other hand, since the confidence region also contains potential other roots of the score, it is not a direct estimate of this distribution.
\end{itemize}

The main idea of the proof is that $\{\|B(\theta, \pi_i)\|^2\}$ are uniformly ordered in case they are evaluated at the true parameter $\theta^*$. The other intuition behind this construction is, similarly to SPS, that as we get farther away from the ``true'' parameter, $\theta^*$, our reference element $\|B(\theta, \pi_0)\|^2$ should  dominate the ordering of $\{\|B(\theta, \pi_i)\|^2\}$: for $i\neq 0$,
$
\|B(\theta, \pi_0)\|^2 \succ_{\nu} \|B(\theta, \pi_i)\|^2,
$
with ``high probability'' as $\theta$ gets ``sufficiently far away'' from $\theta^*$, for example, $\|\theta - \theta^*\|$ is ``large enough''.  However, this property is hard to formulate and prove rigorously, therefore, in Section \ref{sec-exp} we continue with investigating ScoPe experimentally.

\section{EXPERIMENTAL RESULTS}
\label{sec-exp}
Now, we evaluate ScoPe through numerical experiments on simulated data as well as on major stock market indices.
ScoPe is compared with standard asymptotic ellipsoids, residual- and likelihood ratio bootstrap constructions.

Here we focus on GARCH($1$, $1$) processes that constitute a very important special case, as they are by far the most widely used GARCH models in industry \citep{ruppert2011statistics} as well as typical reference models in
empirical comparisons \citep{francq2011garch}. An explanation of this was given by \citet{hansen2005forecast}, who compared the forecasting potential of $330$ volatility models on historical exchange rates and found no statistical evidence that more sophisticated models could outperform GARCH($1$, $1$).

In our first simulated experiment 
the driving noise $\{\varepsilon_t\}$ of the GARCH($1$, $1$) model had logistic distribution with zero mean and scale $\sqrt{3}/\pi$, to ensure unit variance. Thus,
\begin{eqnarray*}
X_t & \!\!\triangleq\!\! & \sigma_t \,\varepsilon_t,\\[5pt]
\sigma^2_t & \!\!\triangleq\!\! & \omega^* + \alpha^* X^2_{t-1}+\beta^* \sigma^2_{t-1},
\end{eqnarray*}
where the true parameter vector was $\theta^* = [\alpha^*, \beta^*, \omega^*]$, where $\alpha^* = 0.44$, $\beta^* = 0.33$ and, since we assumed a system with unit variance, i.e., weak white noise, 
$
\omega^* = 1 - \alpha^* - \beta^* = 0.23.
$
Because of this, it is enough to build a confidence region for $(\alpha^*, \beta^*)$, as they determine $\omega^*$.

\begin{figure}[t]
\centering
      \centering
      \includegraphics[width=.97\linewidth]{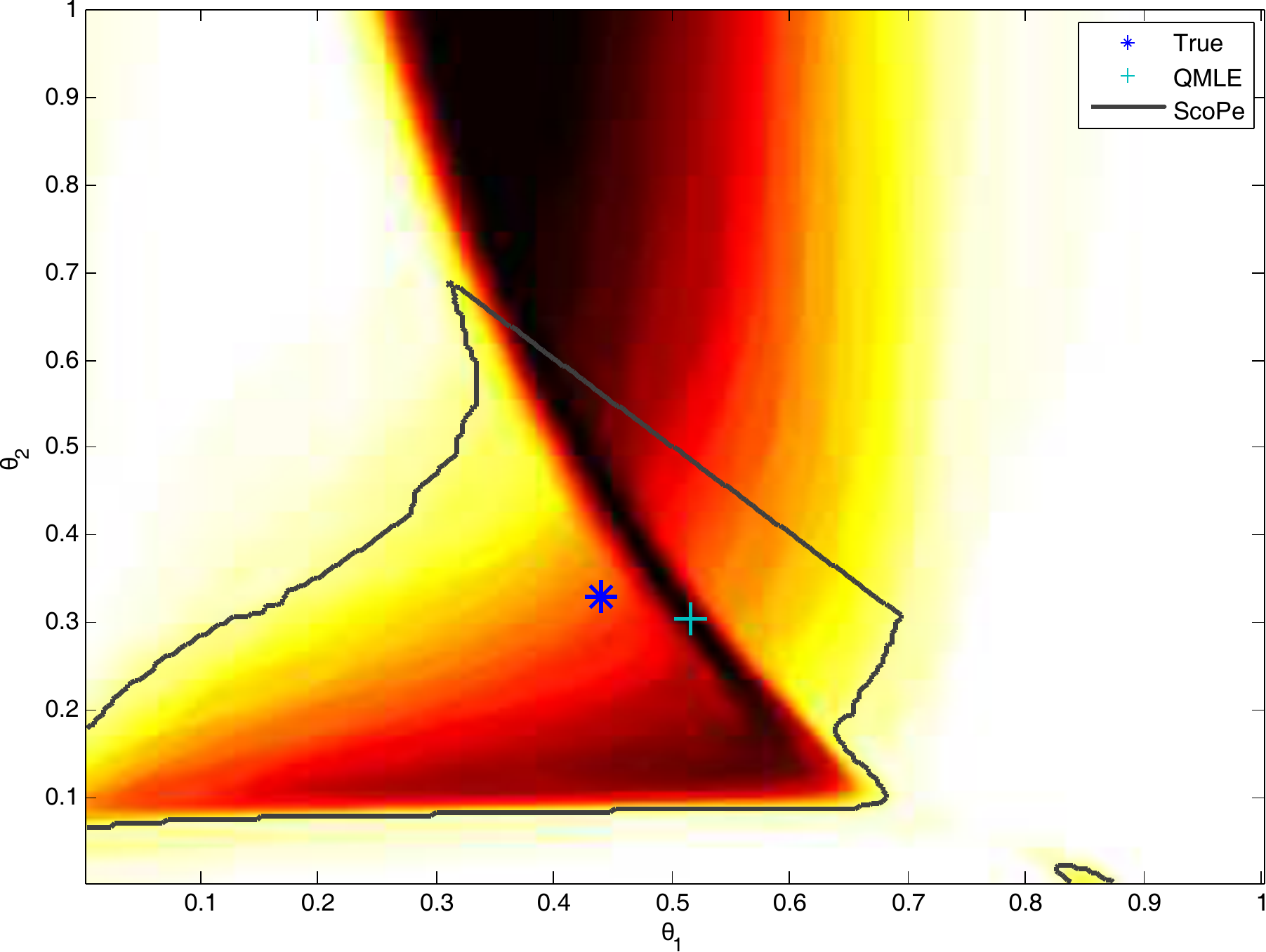}
      \caption{Logistic noise, $n = 100$, $m=100$, $r = 10$; Exact $90$\% ScoPe confidence set for a GARCH$(1,1)$ process and its rank field. Darker color indicates smaller rank.}
      \label{fig:sfig1}
 	  \vspace*{-3mm}
\end{figure}

In order to test the method, $100$ observations were generated. The rank of $\| B_0(\theta)\|^2$, $\mathcal{R}_m(\theta)$, was then calculated for parameters in $[0, 1] \times [0,1]$. The resulting ``rank field'' is shown in Figure \ref{fig:sfig1} with its $90$\,\% confidence region, in which parameters leading to only non-stationary processes were also eliminated, namely, the ones with $\alpha + \beta \geq 1$. 
Note that this region has two connected components.

\begin{figure}[t]
      \centering
      \includegraphics[width=.97\linewidth]{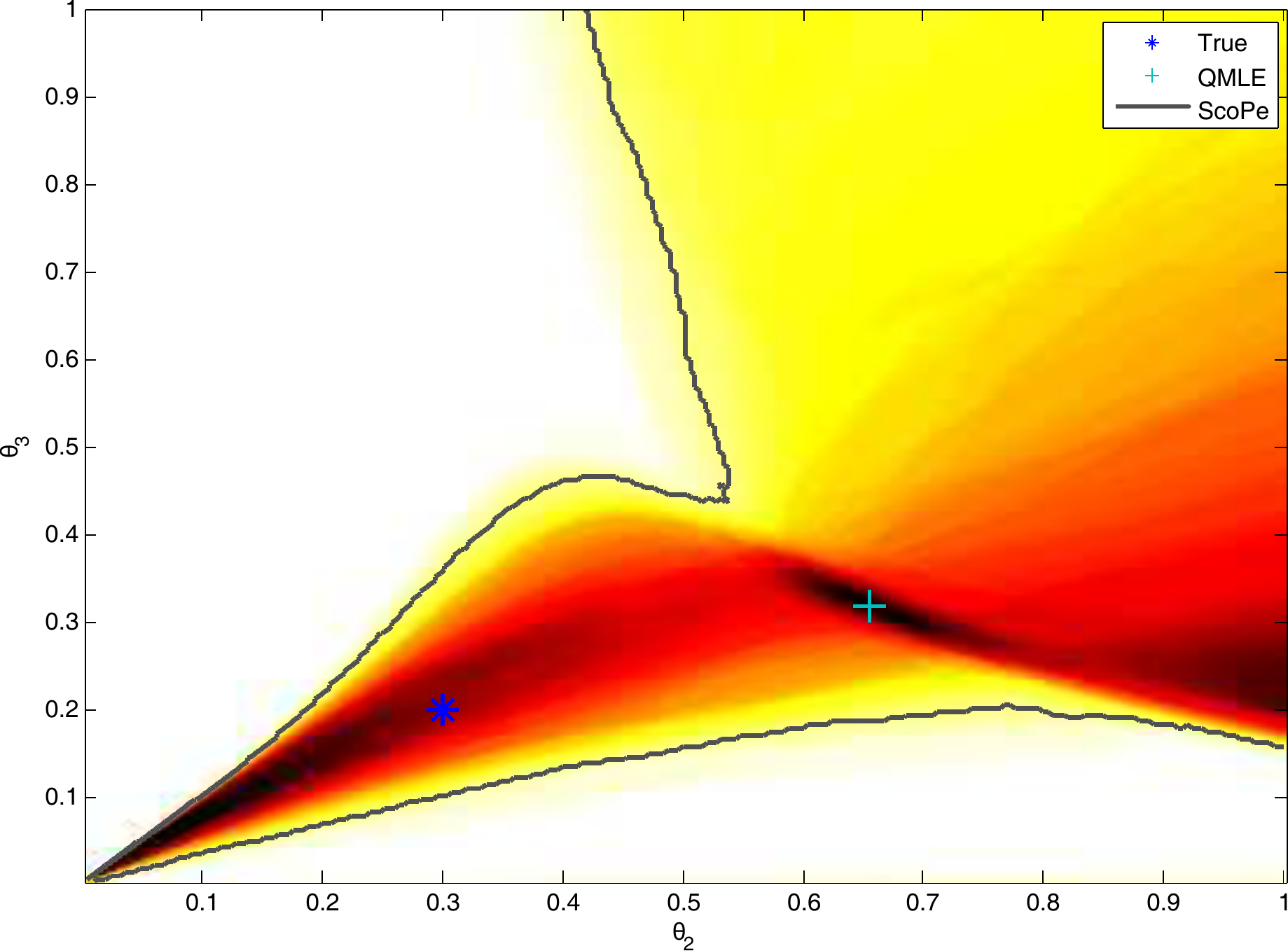}
      \caption{Laplacian noise, $n = 1000$, $m=100$, $r = 10$; The rank of $\|B((\theta), \pi_0)\|^2$ is shown as a function of $(\beta, \omega)$; $\alpha$ is fixed at its QMLE. Darker color indicates smaller rank.}
      \label{fig:sfig2}
		\vspace*{-3mm}
\end{figure}

In our next simulated experiment, illustrated in Figure \ref{fig:sfig2}, the process was generated using Laplacian innovations. Now, the $\omega$ parameter was also estimated ($90$\,\% confidence was targeted), and the true parameter was $\theta^* = [0.44, 0.33, 0.22]$.  As we can observe from the image, the QMLE and the true parameter are situated in different rank-valleys, which explains disconnected confidence regions.

Table \ref{numtable} compares $90$\,\%  confidence sets of the asymptotic approach (\ref{asymptotic ellipsoid}), the residual \citep{pascual2006bootstrap} and (Gaussian) likelihood ratio \citep{luger2012finite} bootstraps and ScoPe. In this experiment Gaussian and Logistic driving noises were applied. Both the empirical coverage of the true parameter and the relative size of the confidence sets compared to the whole class of allowed models (weak white noise assumption) were evaluated by $1000$ Monte Carlo trials. Random noises were generated and it was tested whether $\theta^*$ is in the confidence set (empirical coverage). Next several random parameter values were selected and the relative size of the confidence regions were estimated as the ratio of those parameters which were fallen into the set (relative area). A $1000$ long ``burn in'' simulation was used for initialization, thus assumption \ref{P2} was violated. Nevertheless, ScoPe provided close to exact coverage probabilities combined with relatively small confidence regions.

{\renewcommand{\arraystretch}{1.1}
\begin{center}
\begin{table}[b]
\vspace{1mm}
\small
\caption{Empirical Coverages and Areas on Simulated Data}
\vspace{-1mm}
\begin{center}
\begin{tabular}{lcccc}
\multicolumn{1}{c}{}&\multicolumn{2}{c}{\bf Gaussian Noise}&\multicolumn{2}{c}{\bf Logistic Noise}\\
{\bf Method\hspace{-3mm}} & {\bf Emp.~Cov.\hspace{-2mm}} & {\bf Rel.~Area\hspace{-2mm}} & {\bf Emp.~Cov.\hspace{-2mm}} & {\bf Rel.~Area\hspace{-2mm}}\\
\hline\vspace{-2mm}\\
Asym.Ell. & 0.8656 & 0.4715 & 0.8264 & 0.5446\\ 
Res.Boots. & 0.8567 & 0.7051 & 0.8152 & 0.6139\\ 
LR.Boots. & 0.9655 & 0.6623 & 0.9762 & 0.7681 \\ 
ScoPe& 0.8961 & 0.5324 & 0.9147 & 0.6727  \\ 
\end{tabular}
\end{center}
\label{numtable}
\vspace{6mm}
\small
\caption{Relative Areas on Stock Market Indices (2014)}\label{PerfEvalStock}
\vspace{-2mm}
\begin{center}
\begin{tabular}{lccc}
{\bf Method} & {\bf Nasdaq 100} & {\bf S\&P 500} & {\bf FTSE 100}\\
\hline\vspace{-2mm}\\
Asym.Ell. & 0.3426 & 0.1679 & 0.1535 \\ 
Res.Boots. & 0.3791 & 0.2549 & 0.2850\\ 
LR.Boots. & 0.8150 & 0.7919 & 0.8326\\ 
ScoPe& 0.3801 & 0.2862 & 0.2412\\ 
\end{tabular}
\end{center}
\label{numtable2}
\vspace*{-3mm}
\end{table}
\end{center}}

\vspace{-2mm}
In our last experiment we evaluated the methods on major stock market indices. More precisely, the daily closing prices of Nasdaq 100, S\&P 500 and FTSE 100 were used from the entire period of 2014 (which means 252 observations for each dataset). First, the {\em compound returns} \citep{francq2011garch} were calculated from the data, i.e., for each sequence $\{P_t\}$, the data were transformed by $R_t = \log(P_t/P_{t-1})$. They were then standardized, after which GARCH($1$, $1$) models were estimated using QML.

Only the relative areas of $90$\,\%  confidence regions, approximated using $1000$ Monte Carlo trials, are shown in Table \ref{numtable2} as ``true'' parameters were not available.  The size (but not the shape) of ScoPe confidence sets were about the same as the ones obtained by residual bootstrap, indicating the promising practical applicability of ScoPe, especially, since it has stronger theoretical guarantees than bootstrap.

\vspace{-1mm}
\section{CONCLUSIONS}
\vspace{-1mm}
GARCH processes are widespread models of (conditional) heteroscedasticity, and are archetypically estimated by the QML mehtod, which provides point estimates. Often confidence regions are also needed, but unfortunately the standard approach (based on limiting distributions) fails in case the driving noise is heavy-tailed. Alternative approaches, such as bootstrap methods, may also fail for skewed distributions or require knowledge about the noise terms.

In this paper the ScoPe method was proposed which is based on permuting the score function. At the best of our knowledge, it is the first approach that can construct (i) exact, (ii) non-asymptotic, (iii) distribution-free confidence regions (iv) around the QMLE, (v) without additional assumptions about moments or stationarity. Its exact coverage probability was proved and numerical experiments on simulated as well as stock market data were also presented.
 
\subsubsection*{Acknowledgments}
The work of B.\ Cs.\ Cs\'aji was supported by the Hung.\ Sci.\ Res.\ Fund (OTKA), projects 113038 and 111797, and by the J\'anos Bolyai Research Fellowship, BO\,/\,00683\,/\,12\,/\,6. 

\bibliographystyle{plainnat}
\bibliography{ScoPe-AISTATS-2016}

\end{document}